\documentclass[runningheads]{llncs}
\usepackage[T1]{fontenc}
\usepackage{graphicx}
\usepackage{subcaption}
\usepackage{tabularx} 
\usepackage{amsmath}
\usepackage{algorithm}
\usepackage{algorithmic}
\usepackage{booktabs}
\usepackage{multirow}
\usepackage{enumerate}
\usepackage[hidelinks]{hyperref}
\usepackage{placeins}

\begin{document}
    \title{Protecting Quantum Circuits Through Compiler-Resistant Obfuscation}
    
    \author{
    Pradyun Parayil\inst{1}\thanks{Both authors contributed equally to the work.} \and
    Amal Raj\inst{2}\textsuperscript{\thefootnote} \and
    Vivek Balachandran\inst{2}
    }

    \authorrunning{Pradyun et al.}
    
    \institute{
    Dept. of Electronics and Communication Engineering, Amrita School of Engineering,
    Amrita Vishwa Vidyapeetham, Coimbatore, India, 641112\\
    \email{cb.en.u4cce23067@cb.students.amrita.edu}
    \and
    Singapore Institute of Technology, Singapore\\
    \email{\{amal.raj,vivek.b\}@singaporetech.edu.sg}
    }
    
    \maketitle
    \begin{abstract}
        Quantum circuit obfuscation is becoming increasingly important to prevent theft and reverse engineering of quantum algorithms. As quantum computing advances, the need to protect the intellectual property contained in quantum circuits continues to grow. Existing methods often provide limited defense against structural and statistical analysis or introduce considerable overhead. In this paper, we propose a novel quantum obfuscation method that uses randomized U3 transformations to conceal circuit structure while preserving functionality. We implement and assess our approach on QASM circuits using Qiskit AER, achieving over 93\% semantic accuracy with minimal runtime overhead. The method demonstrates strong resistance to reverse engineering and structural inference, making it a practical and effective approach for quantum software protection.
    \end{abstract}
    
    
    \section{Introduction}        
        Quantum computing leverages the principles of quantum mechanics, such as superposition, entanglement, and quantum interference, to perform computations that are infeasible for classical computers. It is poised to revolutionize fields like cryptography, where it can break conventional encryption schemes, optimization, where it tackles complex combinatorial problems, and quantum simulation, where it models molecular and material properties for drug discovery and materials science \cite{nielsen2010}. Quantum computers are used in research labs, cloud-based quantum platforms, and increasingly in industrial applications to solve specialized problems that require massive parallelism and probabilistic computation \cite{Preskill2018}.
        As quantum software development accelerates, protecting intellectual property (IP) and securing quantum circuits against reverse
        engineering and structural analysis has become a critical concern. Quantum programs, typically expressed in gate-level formats like QASM, reveal their underlying structure, making it easier for adversaries to extract proprietary logic or algorithmic optimizations. Developing robust obfuscation techniques that preserve functionality while concealing the circuit’s logical structure is essential for ensuring the security and commercial viability of quantum software in practical applications.
        
        
        
        \subsection{Proposed Idea}    
            This work introduces a quantum circuit protection technique that resists critical scrutiny and ensures operational functionality. The obfuscation of quantum circuit topology is accomplished by applying random \texttt{U3} gate transformations. The main idea is to perform a full circuit transformation by enclosing it with a random basis transformation and its dual. Each internal gate is transformed as well with this same transformation, thereby masking their identity while maintaining functionality. All the gates in the circuit are transformed through single-qubit \texttt{U3} matrices, which hide the circuit’s inner workings without affecting its primary function.
            
\begin{figure}[htbp]
              \centering
              \includegraphics[width=1\linewidth]{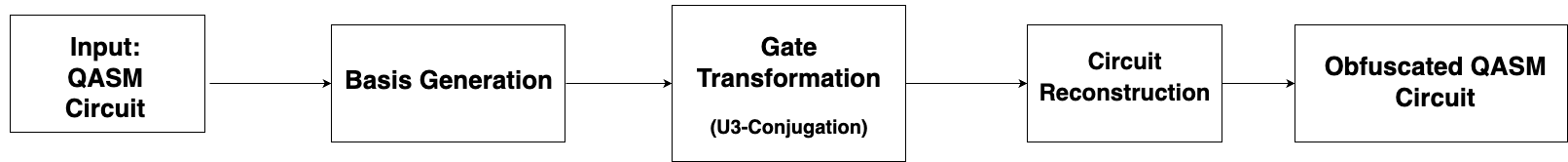}
              \caption{Workflow of the proposed quantum circuit obfuscation using U3-conjugation. The process begins with a QASM circuit, applies a randomized basis transformation, transforms each gate, and restores the original basis to produce an obfuscated circuit with preserved functionality.}
              \label{fig:workflow}
            \end{figure}

            This method makes sure the circuit's output stays the same because the changes cancel each other out. Traditional ways of doing obfuscation add more gates or make the circuit hard to understand, but our method adds only negligible extra overhead. This keeps the circuit quick and efficient on quantum hardware \cite{Preskill2018}. It fits right in with any quantum program in QASM/Qiskit format, does not require modifications to the original algorithm, and is simple to deploy across different quantum setups. Due to these merits, our method is a flexible and practical approach for safeguarding quantum software in real-world scenarios. The overall workflow is illustrated in Fig. \ref{fig:workflow}.
            
        \subsection*{Paper Organization}
            The rest of this paper is structured as follows. Section \ref{sec:background} provides background on quantum computing, including qubits, quantum gates, compilation, and measurement operations. It also reviews existing work on quantum circuit obfuscation, highlighting current techniques and their limitations. Section \ref{sec:methodology} outlines the proposed methodology and presents our obfuscation technique. Section \ref{sec:application} illustrates the methodology through the obfuscation of a QAOA circuit. Section \ref{sec:metrics} presents evaluation metrics and results, demonstrating the effectiveness of the proposed technique.
            Section \ref{sec:security} details the security guarantees provided by our approach, formalizing the threat model and the protections achieved. Section \ref{sec:limitations} discusses the limitations of our approach and proposes directions for future research. Section \ref{sec:conclusion} concludes the paper by summarizing the key contributions and their implications for secure quantum software development.

    \section{Background}\label{sec:background}
        \subsection{Quantum Computing Fundamentals}
            Quantum computing diverges from classical computing by employing quantum bits, or qubits, which leverage quantum mechanics to process information. Unlike classical bits, restricted to a binary state of 0 or 1, qubits can exist in a superposition, embodying multiple states simultaneously. This capability enables quantum computers to perform parallel computations, vastly exceeding classical systems for certain tasks. Additionally, qubits can be entangled, meaning the state of one qubit is directly tied to another, allowing instant correlations across distances. Such properties enable quantum algorithms to tackle complex problems in cryptography, optimization, and beyond with remarkable efficiency \cite{nielsen2010}. Quantum gates, the core of quantum computation, manipulate qubit states through reversible operations, represented as unitary matrices. For instance, the Hadamard gate creates superposition, while the controlled-NOT gate fosters entanglement, enabling intricate quantum interactions \cite{rieffel2011}.
            
            Quantum circuits, composed of sequences of quantum gates, orchestrate computations by evolving qubit states toward a desired outcome. A single measurement yields one classical outcome, whereas repeated executions of the circuit generate a probability distribution over possible results. Quantum Assembly Language (QASM), particularly its evolution from \texttt{OPENQASM 2.0} to \texttt{OPENQASM 3.0}, standardizes circuit descriptions, supporting classical control flow and dynamic operations \cite{Cross2017}. This facilitates circuit portability across quantum platforms but also raises concerns about protecting intellectual property embedded in these circuits. Compilation, a vital step for executing circuits on physical hardware, translates high-level designs into hardware-specific instructions. Tools like Qiskit’s transpiler optimize gate sequences and map qubits to physical hardware, minimizing circuit depth to enhance performance on Noisy Intermediate-Scale Quantum (NISQ) devices. Yet, this process exposes circuits to potential reverse-engineering, as adversaries could analyze gate structures to extract proprietary algorithms \cite{obfusqate2025}.
        
        \subsection{Related Works}
            Several approaches have been proposed to secure quantum circuits through obfuscation, each with distinct trade-offs. Suresh et al. introduced a method that inserts a dummy gate at a specific circuit location to obscure its functionality, intentionally disrupting results during execution \cite{suresh2021}. Post-compilation, this gate is removed to restore the original computation. To prevent optimization, barriers are placed around the dummy gate, and additional barriers are inserted around other gates to mask its identity from adversaries. However, this reliance on barriers compromises circuit optimization, increasing depth and degrading performance on NISQ hardware \cite{suresh2021}. Similarly, Rehman, Langford and Liu proposed an obfuscation technique that modifies gate parameters using secret keys to conceal circuit logic \cite{rehman2025}. While effective in hiding structure, this method requires significant classical processing for key management, complicating deployment and scalability, particularly for resource-constrained quantum systems. Another approach by Bartake et al. uses redundant gate insertion to obscure the intent of the circuit. This improves security, but the added gates significantly increase the depth of the circuit, making it more prone to errors on NISQ devices and limiting practical applicability \cite{obfusqate2025}. These methods highlight the challenge of balancing security with performance, as increased complexity often undermines execution fidelity on current quantum hardware \cite{Preskill2018,suresh2021}.

    
        \section{Methodology}\label{sec:methodology}
             Our quantum circuit obfuscation technique leverages basis transformations via U3 conjugation to obscure gate identities while preserving functional correctness. For clarity, the formulation presented in this paper adopts the weakest variant of the obfuscation framework, where a single random basis transformation is applied uniformly across the circuit. This setting intentionally models the worst case for security analysis and served as the baseline design during development. A strictly stronger variant assigns an independently sampled U3 basis to every gate, which removes global structural correlations and drives an adversary’s inference probability to negligible values. A detailed discussion of the resulting security guarantees is presented in Section~\ref{sec:security}.
            
            The technique proceeds in four phases: input parsing, basis generation, gate transformation, and circuit reconstruction. Together, these stages preserve QASM compatibility, maintain semantic equivalence, and enhance resistance to reverse engineering.
    
        \subsection{QASM Input Parsing}
            The first step in our quantum circuit obfuscation process is to accurately interpret the user-provided quantum circuit, which is specified in Quantum Assembly Language (QASM). QASM serves as a standardized, platform-independent format for describing quantum circuits, including gate operations, qubit allocations, and measurement instructions \cite{Cross2017}. To ensure compatibility with a wide range of quantum circuits, our system is designed to parse input strings written in either \texttt{OPENQASM 2.0} or the more advanced \texttt{OPENQASM 3.0} specification, accommodating both legacy and modern quantum programming workflows.
    
    
            The input QASM string is first trimmed to remove any extra whitespace. Based on its prefix, the QASM version is identified: Qiskit uses \textit{qiskit.qasm2.loads()} for OpenQASM 2.0 and \textit{qiskit.qasm3.loads()} for OpenQASM 3.0, which supports advanced features like control flow and custom gate definitions \cite{Cross2017,qasm3}. If the version is unrecognized, an error is raised. This process produces a QuantumCircuit object with the circuit’s gates, qubits, and measurements.
    
        \subsection{Basis Generation}
            The core of our obfuscation relies on a randomized basis transformation applied across all qubits. We generate random parameters ($\theta \in [0,\pi]$, $\phi \in [0,2\pi]$, $\lambda \in [0,2\pi]$) for a U3 gate using a uniform random distribution. The U3 gate is a universal single-qubit gate  that can represent any unitary operation by tuning its parameters, making it well-suited for implementing a wide range of basis transformations. Randomization ensures that each circuit instantiation produces a unique basis, enhancing obfuscation by preventing predictable patterns that could be exploited in reverse engineering. As shown in Equation \ref{eq:u3_gate}, the gate's matrix representation is defined by these parameters.
    
            \begin{equation}
            \label{eq:u3_gate}
                U3(\theta, \phi, \lambda) =
                \begin{bmatrix}
                    \cos\left(\frac{\theta}{2}\right) & -e^{i\lambda} \sin\left(\frac{\theta}{2}\right) \\
                    e^{i\phi} \sin\left(\frac{\theta}{2}\right) & e^{i(\phi + \lambda)} \cos\left(\frac{\theta}{2}\right)
                \end{bmatrix}
            \end{equation}
    
        \subsection{Gate Transformation}
            To obfuscate a quantum circuit, each gate $G$ in the original circuit is transformed using unitary conjugation, expressed as $U^{\dagger} \cdot G \cdot U$, where $U$ is the unitary matrix of a \texttt{U3} gate with randomly chosen parameters $(\theta, \phi, \lambda)$. Measurements and resets are left unchanged because the obfuscation method applies only to unitary operations. Since measurement is non-unitary, conjugating it with $U^{\dagger}$ and $U$ would rotate the measurement basis and alter the circuit’s behavior. For circuits containing mid-circuit measurements, we simply treat the computation as a sequence of unitary segments separated by measurement boundaries and obfuscate each unitary segment independently.
            
            Using Qiskit’s \texttt{Operator} class, the gate $G$ is first converted into its matrix representation to facilitate this transformation. For multi-qubit gates, the basis matrix $U$ is constructed by computing the Kronecker product of single-qubit \texttt{U3} matrices for each qubit. The Kronecker product, a tensor operation that combines matrices to form a larger matrix, ensures that the transformation scales appropriately for systems with multiple qubits, preserving the gate’s functionality across the circuit’s qubit register \cite{Laub2005}. The resulting matrix $U^{\dagger} \cdot G \cdot U$ is then encapsulated as a \texttt{UnitaryGate} with a unique label, such as \texttt{Obf\_H\_0} for a Hadamard gate, effectively concealing the original gate’s identity. This process is applied to all gates, as detailed in Algorithm \ref{alg:gate-transform}.
    
            \begin{algorithm}
                \caption{Gate Transformation via U3-Conjugation}
                \label{alg:gate-transform}
                \begin{algorithmic}[1]
                    \REQUIRE Gate $G$, \texttt{qubits}, \texttt{U3} parameters $(\theta, \phi, \lambda)$
                    \STATE Convert $G$ to matrix using \texttt{Operator(G)}
                    \STATE Compute \( U = U3(\theta, \phi, \lambda)^{\otimes n} \) for \( n \) qubits
                    \STATE Compute \( U^{\dagger} = U3(-\theta, -\lambda, -\phi)^{\otimes n} \)
                    \STATE Compute transformed matrix \( U^{\dagger} \cdot G \cdot U \)
                    \STATE Encapsulate as \texttt{UnitaryGate} with label \texttt{Obf\_G\_id}
                    \STATE Replace original gate $G$ with the transformed \texttt{UnitaryGate} in the circuit
                \end{algorithmic}
            \end{algorithm}
            
            \begin{figure}[h]
                \centering
                \includegraphics[width=0.5\linewidth]{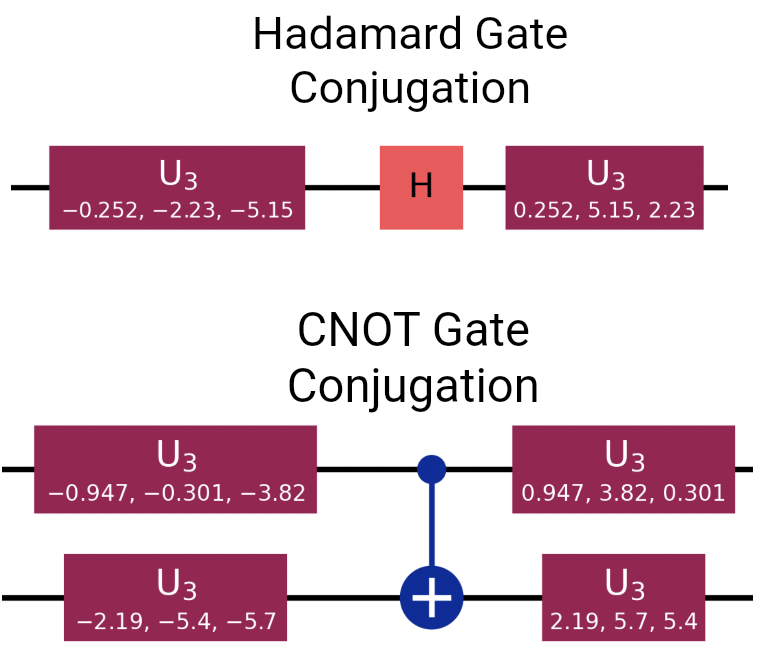}
                \caption{Diagram of the conjugation process $U^{\dagger} \cdot G \cdot U$ for a single-qubit gate (Hadamard) and a multi-qubit gate (CNOT). This figure illustrates the underlying matrix operations and does not imply additional quantum gates or circuit depth. The result of this matrix multiplication ($U^{\dagger} \cdot G \cdot U$) is encapsulated as a single, opaque \texttt{UnitaryGate} object, thereby incurring no space overhead in the quantum circuit.}
                \label{fig:conjugation}
            \end{figure}
            
        \subsection{Basis Transformation and Obfuscation}
            For a circuit with $n$ qubits, a global basis transformation $U = U3(\theta, \phi, \lambda)^{\otimes n}$ is applied to all qubits at the circuit’s start, with its inverse $U^{\dagger} = U3(-\theta,-\lambda,-\phi)^{\otimes n}$ applied at the end. The parameters $\theta$, $\phi$, and $\lambda$ are randomly sampled from a uniform distribution to maximize obfuscation. For each gate $G$, a local basis transformation is applied, replacing $G$ with its conjugated form $G' = U3(-\theta, -\lambda, -\phi) \cdot G \cdot U3(\theta, \phi, \lambda)$, which encapsulates the original gate in a new basis while preserving its functionality. This transformation is implemented using Qiskit’s circuit manipulation tools, where each gate is directly replaced with its transformed version $U^{\dagger} G U$. To ensure the circuit’s original functionality, the inverse basis transformation $U^{\dagger}$ is applied across all qubits as a single instruction labeled \texttt{InvBasis} after all gates are processed. Barriers from the original circuit are retained to maintain structural hints for hardware compatibility. The reconstructed circuit is then transpiled and executed using the Qiskit AER simulator, producing results statistically equivalent to the original circuit. Fig. \ref{fig:conjugation} illustrates this process with examples of Hadamard and CNOT gate transformations. The implementation code is available in the GitHub repository in \cite{basistransformation2025}.
    
        \subsection{Preservation of Functionality: Hadamard and CNOT Gate Examples}
            To illustrate how our obfuscation technique preserves the functionality of quantum circuits, we examine the transformation of two fundamental quantum gates: the single-qubit Hadamard ($H$) gate and the two-qubit controlled-NOT (CNOT) gate. These gates are critical building blocks in quantum algorithms, with the $H$ gate creating superposition and the CNOT gate introducing entanglement \cite{nielsen2010}. By applying our basis conjugation approach to each gate individually, we demonstrate that their computational behavior remains unchanged despite structural obfuscation.
    
            For the Hadamard gate, represented by the $2 \times 2$ unitary matrix $H = \frac{1}{\sqrt{2}} \begin{bmatrix} 1 & 1 \\ 1 & -1 \end{bmatrix}$, we apply a basis transformation using a \texttt{U3} gate with randomly chosen parameters $(\theta, \phi, \lambda)$. The transformation is defined as $U = U3(\theta, \phi, \lambda)$, and the obfuscated gate is computed as:
    
            \begin{equation}
                H_{\text{obf}} = U^{\dagger} \cdot H \cdot U
            \end{equation}
            To verify preservation of functionality, we apply the inverse transformation:
            
            \begin{equation}
                \begin{aligned}
                    U \cdot H_{\text{obf}} \cdot U^{\dagger} = U \cdot      (U^{\dagger} \cdot H \cdot U) \cdot U^{\dagger} = (U        U^{\dagger}) \cdot H \cdot (U U^{\dagger}) = H
                \end{aligned}
            \end{equation}
            
\begin{figure}[H]
                \centering
                \includegraphics[width=1\linewidth]{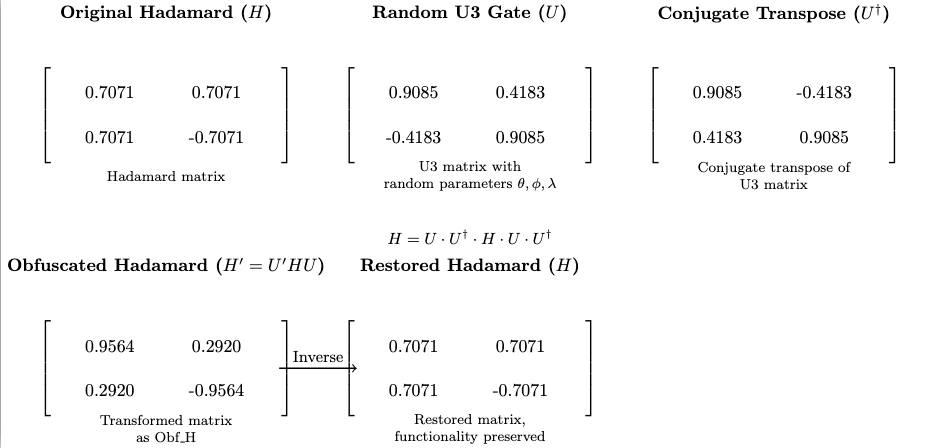}
                \caption{Matrix-level transformation of a Hadamard gate using \texttt{U3} conjugation. The functionality remains unchanged due to the inverse transformation restoring the original gate.}
                \label{fig:hadamard-matrix-conjugation}
            \end{figure}

            This confirms that the obfuscated Hadamard gate, when wrapped by the basis transformation and its inverse, behaves identically to the original $H$ gate, preserving its ability to create superposition. The matrix-level transformation of the Hadamard gate is shown in Fig. \ref{fig:hadamard-matrix-conjugation}.

            For the two-qubit CNOT gate, represented by a $4 \times 4$ unitary matrix, we use a basis transformation $U = U3(\theta, \phi, \lambda)^{\otimes 2}$, formed by the Kronecker product of two single-qubit \texttt{U3} matrices to scale the transformation across both qubits \cite{Laub2005}. The obfuscated CNOT gate is:
            \begin{equation}
                \text{CNOT}_{\text{obf}} = U^{\dagger} \cdot \text{CNOT} \cdot U
            \end{equation}
            Applying the inverse transformation yields:
            \begin{equation}
                \begin{aligned}
                    U \cdot \text{CNOT}_{\text{obf}} \cdot U^{\dagger} &= U \cdot (U^{\dagger} \cdot \text{CNOT} \cdot U) \cdot U^{\dagger} \\
                    &= (U U^{\dagger}) \cdot \text{CNOT} \cdot (U U^{\dagger}) \\
                    &= \text{CNOT}
                \end{aligned}
            \end{equation}
    
\begin{figure}[H]
                \centering
                \includegraphics[width=0.9\linewidth]{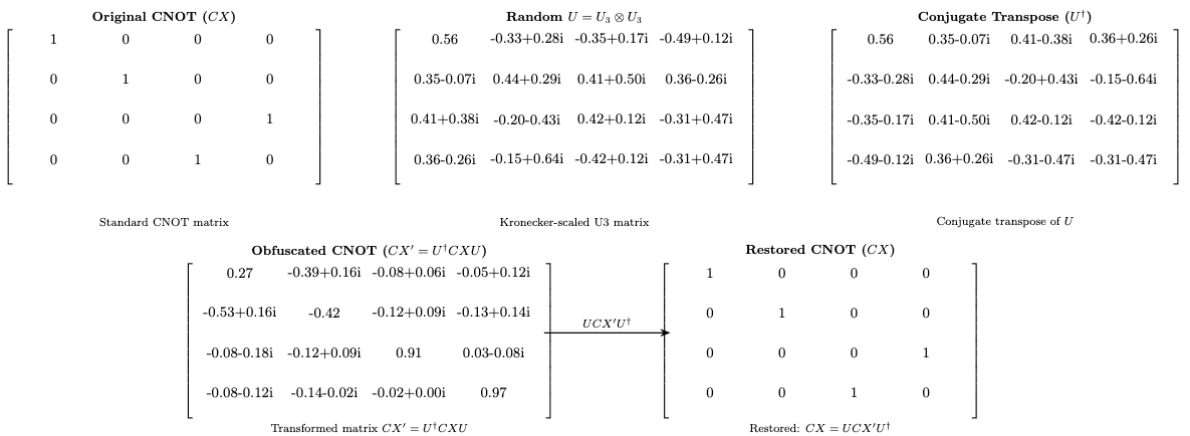}
                \caption{Matrix-level transformation of a CNOT gate using \texttt{U3} conjugation, with the Kronecker product scaling the basis for two qubits.}
                \label{fig:cnot-matrix-conjugation}
            \end{figure}

            This demonstrates that the CNOT gate’s entanglement-generating functionality is fully preserved, as the basis transformations cancel out, leaving the original gate’s behavior intact. This process, applied to individual gates, ensures that the circuit’s overall semantics remain unchanged, as the global basis shift and its inverse cancel out during execution. The corresponding matrix-level transformation for the CNOT gate is shown in Fig.~\ref{fig:cnot-matrix-conjugation}.
    
            
    
    \section{Case Study: Quantum Approximate Optimization Algorithm (QAOA)}\label{sec:application}
        \subsection{Overview of QAOA}
        
        The Quantum Approximate Optimization Algorithm (QAOA) is a hybrid quantum-classical algorithm designed for solving combinatorial optimization problems such as MaxCut and Max3SAT \cite{farhi2014}. It combines parameterized quantum circuits with classical optimization to approximate the solution to a given cost function. QAOA is particularly relevant due to its practical applicability on near-term quantum devices and its structured circuit layout, making it a suitable candidate for testing circuit obfuscation techniques.
        
        In this case study, we focus on a QAOA circuit targeting the MaxCut problem on a 5-node graph. The circuit involves alternating applications of the problem (cost) unitary and a mixing unitary, parameterized by angles $\gamma$ and $\beta$, respectively. We implement this circuit in \texttt{OPENQASM 2.0}, simulate its behavior before and after obfuscation, and evaluate functional correctness and performance overhead.
        
        \subsection{Circuit Description}  
\begin{figure}[h]
            \centering
            \includegraphics[width=0.6\linewidth]{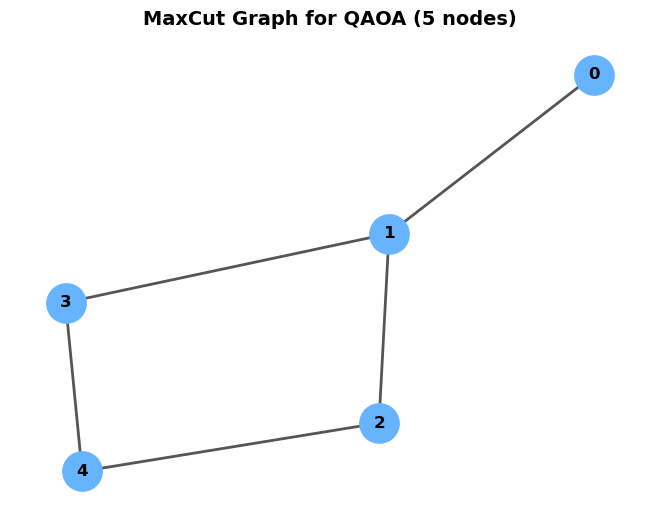}
            \caption{MaxCut graph corresponding to the 5-node QAOA circuit.}
            \label{fig:graph_qaoa}
        \end{figure}

        The QAOA circuit is constructed with five qubits representing the five graph nodes in the MaxCut problem, where the edges are $(0,1), (1,2), (1,3), (3,4), (2,4)$ as shown in Fig.~\ref{fig:graph_qaoa}. The initial state is prepared using Hadamard gates to create a uniform superposition over all possible bitstrings. The cost unitary is implemented using two-qubit $R_{ZZ}$ rotations (decomposed into \texttt{CX--RZ--CX}) corresponding to the edges in the graph. Each edge contributes a term in the cost Hamiltonian, and the circuit applies these terms with an optimized parameter $\gamma$. This is followed by a layer of $R_X$ gates applied to each qubit, representing the mixer Hamiltonian and controlled by the parameter $\beta$.  

        \FloatBarrier

\begin{figure}[h]  
          \centering  
          \includegraphics[width=1\linewidth]{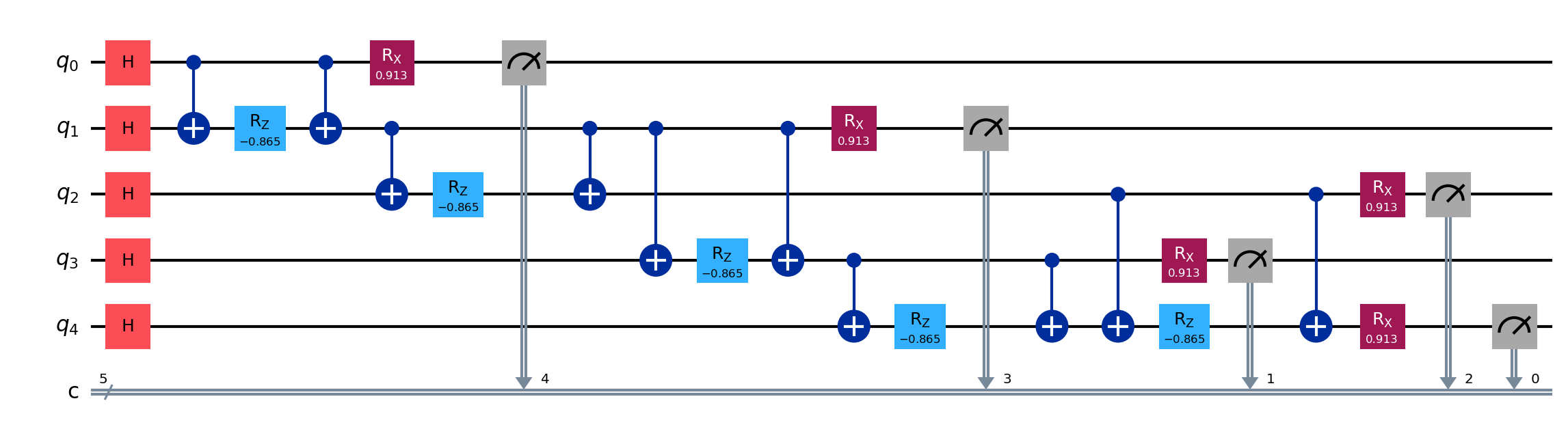}  
          \caption{QAOA circuit diagram for MaxCut on a 5-node graph, depth $p=1$ with optimized parameters.}  
          \label{fig:qaoa_circuit}  
        \end{figure}

        In this implementation, the QAOA depth is $p=1$, meaning one alternating layer of cost and mixer unitaries is applied. The parameters were optimized via classical simulation to $\gamma \approx 0.865$ and $\beta \approx 0.457$, producing a strong probability concentration on the optimal MaxCut solutions.
        The corresponding QAOA circuit structure for this problem is illustrated in Fig.~\ref{fig:qaoa_circuit}. The two optimal MaxCut bitstrings for this graph are \texttt{01001} and \texttt{10110}.

        \subsection{Circuit Obfuscation}
\begin{figure}[h]
              \centering
              \includegraphics[width=0.8\linewidth]{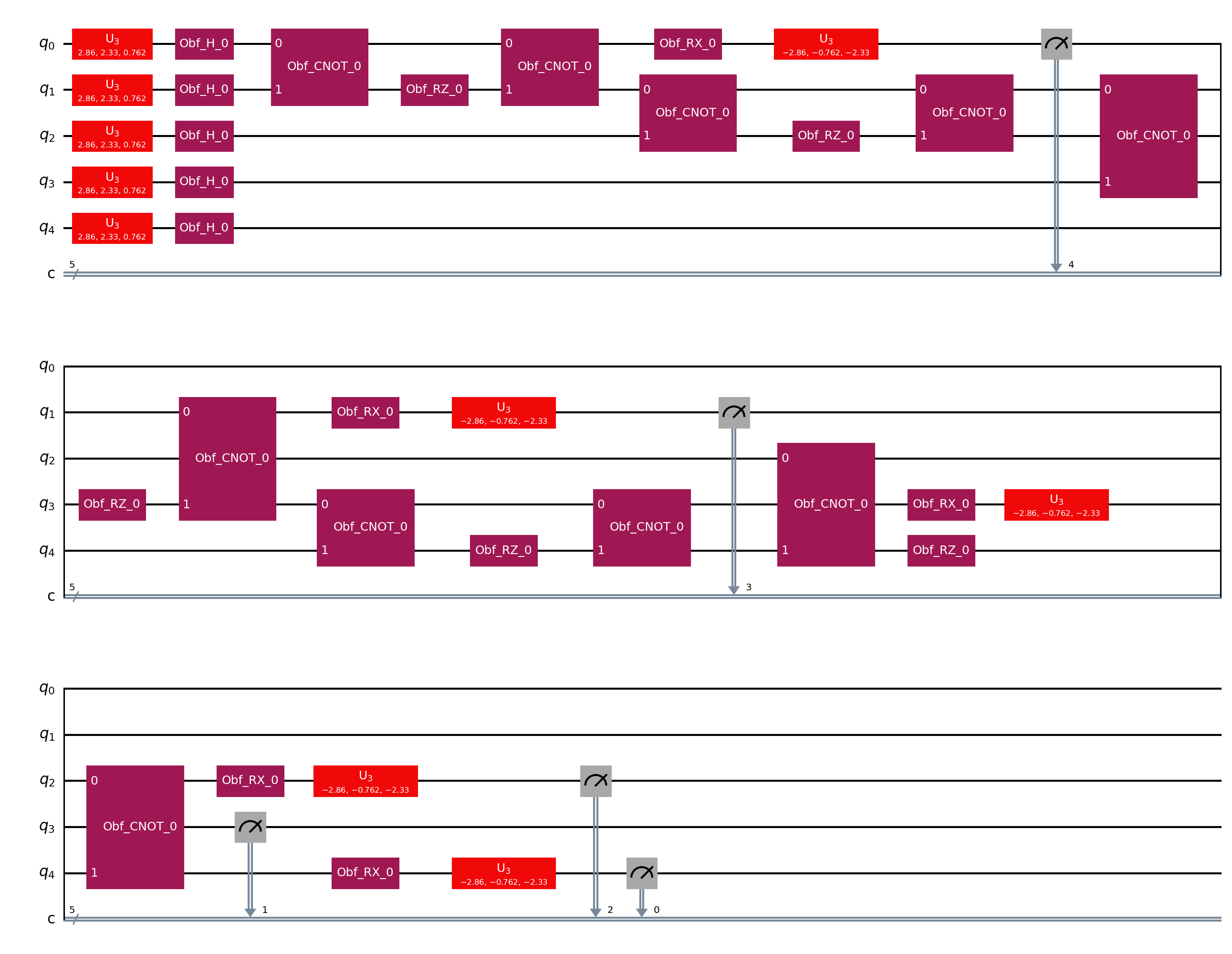}
              \caption{Obfuscated QAOA circuit diagram, showing the transformed gate structure.}
              \label{fig:obfuscated_circuit}
            \end{figure}

            To apply our obfuscation methodology (Section \ref{sec:methodology}) to the QAOA circuit, we generated a random single-qubit basis transformation using $U3(\theta, \phi, \lambda)$ with parameters 
            $\theta = 2.86$, $\phi = 2.33$, and $\lambda = 0.762$. This unitary was then extended to the five-qubit system as a global basis 
            $U = U3(2.86, 2.33, 0.762)^{\otimes 5}$ applied across all qubits at the beginning of the circuit, with its inverse 
            $U^{\dagger} = U3(-2.86, -0.762, -2.33)^{\otimes 5}$ appended at the end to preserve the circuit’s overall functionality. Each gate $G$ in the QAOA layers was obfuscated using the conjugation 
            $G' = U^{\dagger} G U$ and encapsulated as an opaque \texttt{UnitaryGate} with a unique label (e.g., \texttt{Obf\_H\_0}). The transformed structure of the obfuscated QAOA circuit is illustrated in Fig.~\ref{fig:obfuscated_circuit}. 
            
\begin{figure}[h]
              \centering
              \includegraphics[width=1\linewidth]{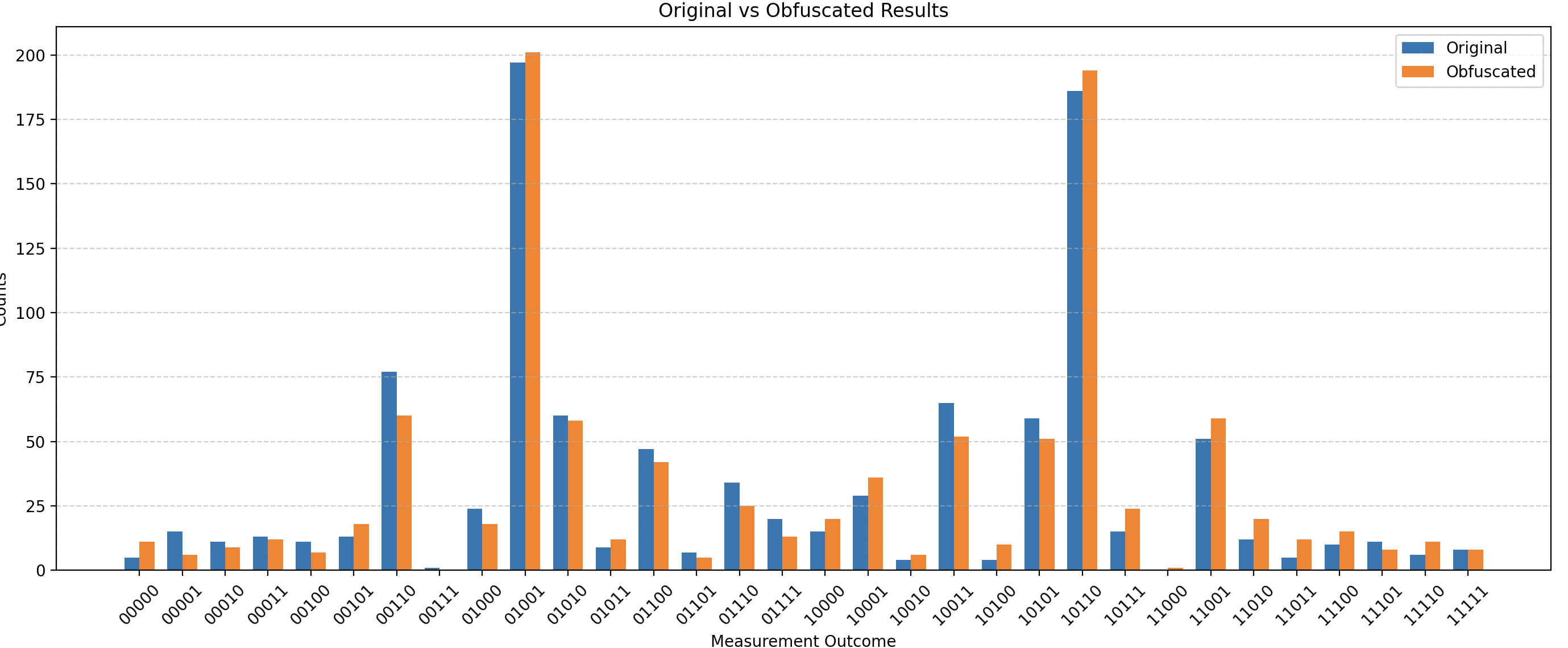}
              \caption{Histogram of measurement outcomes for the original and obfuscated QAOA circuits. Optimal solutions are highlighted.}
              \label{fig:original_results}
            \end{figure}

            The resulting obfuscated circuit is functionally equivalent to the original but structurally unreadable, providing resistance against reverse engineering. Both the original and obfuscated circuits were then tested under identical simulation settings. The measurement outcomes for both the original and obfuscated QAOA circuits are shown in Fig. \ref{fig:original_results}. 
            
            \FloatBarrier

        \subsection{Testing Process}  
            To evaluate the obfuscation technique, both the original and obfuscated QAOA circuits were executed using Qiskit’s \texttt{AerSimulator} (version 0.36.0). Each configuration was run with 1,024 shots to ensure sufficient sampling. The output distributions were recorded and plotted using histograms. Execution times were also measured to detect any latency introduced by obfuscation.  
        
            The comparison focused on:  
            \begin{enumerate}[i.]  
                \item \textbf{Semantic accuracy} --- probability assigned to each bitstring in the obfuscated circuit compared to the original.  
                \item \textbf{Total Variation Distance (TVD)} --- quantifying statistical difference between distributions.  
                \item \textbf{Functional equivalence} --- whether optimal solutions remain present with comparable probability.  
                \item \textbf{Execution time} --- runtime overhead introduced by the obfuscation process
            \end{enumerate}  
        
        \subsection{Results}
            Both circuits produce highly similar output distributions, with only minor statistical variation due to sampling noise. The semantic accuracy of the obfuscated circuit reached \textbf{93.30\%}, and the TVD was calculated to be \textbf{0.0840}, indicating strong fidelity to the original behavior. The two optimal MaxCut bitstrings (\texttt{01001} and \texttt{10110}) emerged as the two highest-probability peaks in both runs, dominating the output distributions with comparable counts. Execution time measurements showed negligible overhead introduced by the obfuscation process.  
        
            These results confirm that the obfuscation strategy preserves the computational intent and functional output of the original circuit, with the chosen optimized QAOA parameters successfully amplifying the probability of the optimal solutions.
                
        \subsection{Implementation Setup}
            The simulations were conducted on a standard Macbook Air M1 with 16 GB RAM and MacOS 18.6. To provide a comprehensive evaluation of our obfuscation technique,  Quantum Approximate Optimization Algorithm (QAOA) for combinatorial optimization was tested alongside nine other quantum circuits, representing a diverse set of quantum algorithms: Shor’s algorithm for integer factorization \cite{Shor1997},  the Bernstein-Vazirani (BV) algorithm for learning hidden linear functions \cite{Bernstein1997}, Grover’s algorithm for unstructured search \cite{Grover1996}, the Variational Quantum Eigensolver (VQE) for quantum chemistry simulations \cite{Peruzzo2014}, the Phase Kickback Circuit \cite{Cleve_1998}, the Deutsch-Jozsa (DJ) algorithm for classifying balanced or constant functions \cite{Deutsch1992}, the Quantum Fourier Transform (QFT) for periodicity analysis \cite{qft}, and Simon’s algorithm for finding periods in functions \cite{Simon1997}. These algorithms, spanning applications from cryptography to quantum simulations, provide a robust testbed for assessing the obfuscation method’s versatility.
    
            Each circuit, including QAOA, was executed 100 times with 1,024 shots per execution under noise-free conditions using Qiskit’s AerSimulator. The Toffoli circuit used three qubits, while the other algorithms ranged from 4 to 12 qubits, depending on their complexity.
       
    \section{Evaluation Metrics}\label{sec:metrics}
        These circuits were executed under noise-free conditions using Qiskit’s AerSimulator to ensure consistent and reproducible results. The evaluation focuses on three primary metrics: \textit{semantic accuracy}, \textit{Total Variation Distance (TVD)}, and \textit{execution time}, which together assess functional fidelity, statistical similarity, and computational overhead.

        \subsection*{Semantic Accuracy}  
            Semantic accuracy counts the overlapping probability mass between the two distributions, expressed as a percentage of the total probability in the original circuit's output. It quantifies whether the obfuscated circuit preserves the functional behavior of the original circuit under realistic, finite-shot quantum execution. This metric was used by Bartake et al. to quantify preservation of circuit behavior after their circuit-level transformations \cite{obfusqate2025}. Given two measurement distributions \texttt{original} and \texttt{obfuscated}, the semantic accuracy is calculated as:
            
            \begin{equation}  
                \text{Semantic~Accuracy} = \frac{\sum\limits_{x \in X} \text{min} (\text{original}[x], \text{obfuscated}[x])}{\sum\limits_{x \in X} \text{original[x]}} \times 100\%
            \end{equation} 
            where $X$ is the union of all bitstrings observed in either distribution.

            Since quantum measurements are inherently probabilistic, two runs of the same circuit will exhibit slight differences in their empirical output distributions unless an infinite number of shots is used. Semantic accuracy captures this agreement by computing the fraction of output bitstrings whose measured probabilities in the obfuscated circuit fall within the statistical variation of those from the original circuit.

            Values slightly below 100\% are therefore expected even for functionally identical circuits and do not indicate any semantic deviation introduced by the obfuscation. Instead, they reflect natural run-to-run statistical fluctuations. In the idealized infinite-shot limit, the empirical distributions of equivalent circuits would converge exactly, and semantic accuracy would approach 100\%. High semantic accuracy in our experiments (typically above 90\%) demonstrates that the obfuscated circuits remain consistent with the intended computation within the limits of finite-shot quantum sampling.
            
        \subsection*{Total Variation Distance (TVD)}
            Total Variation Distance (TVD) is a statistical measure of the difference between two probability distributions \cite{tvd}. In this work, TVD is used to quantify the effect of gate additions introduced during obfuscation.{TVD is also a standard and widely used metric for comparing output distributions of quantum circuits, and it has been adopted in several prior works on quantum obfuscation \cite{wang2025,Liu2025}. It is computed as the sum of the absolute differences between the measurement counts of the original and obfuscated circuits, normalized by the total number of shots, as given by:
            
            \begin{equation}
                \begin{aligned}
                    \mathrm{TVD} = \frac{\sum\limits_{i} \left| x_{i,\mathrm{obfus}} - x_{i,\mathrm{orig}} \right|}{2 \times \text{Number of shots}}
                \end{aligned}
            \end{equation}
            where $x_{i,\mathrm{orig}}$ and $x_{i,\mathrm{obfus}}$ denote the counts of the $i^{\text{th}}$ measurement outcome for the original and obfuscated circuits, respectively. A TVD of $0$ means the two output distributions are exactly the same, showing no statistical difference at all, while a TVD of $1$ means the distributions are completely distinct with no overlap, representing the maximum possible statistical difference.  
           
     \subsection*{Execution Time}
        Execution time refers to the duration required by the simulator to execute the circuit for a fixed number of shots (1024 in our case), excluding compilation time. It serves as a practical measure of the computational overhead introduced by circuit-level transformations, and its use is consistent with prior work in both quantum and classical settings. Studies on quantum program performance explicitly analyze execution time as a key metric to evaluate runtime and performance overhead in simulators and hardware \cite{ma2025}, while recent work in binary obfuscation likewise evaluates execution time overhead to quantify the performance cost of structural code transformations \cite{zhang2025}.} Our results show that the overhead is minimal (typically $< 1$ ms), making the approach practical for real-world quantum workloads.

        \subsection{Obfuscation's Compiler Resistance: A Key Advantage}\label{sec:compiler_resistance}
\begin{figure}[h]
                \centering
                \includegraphics[width=1\linewidth]{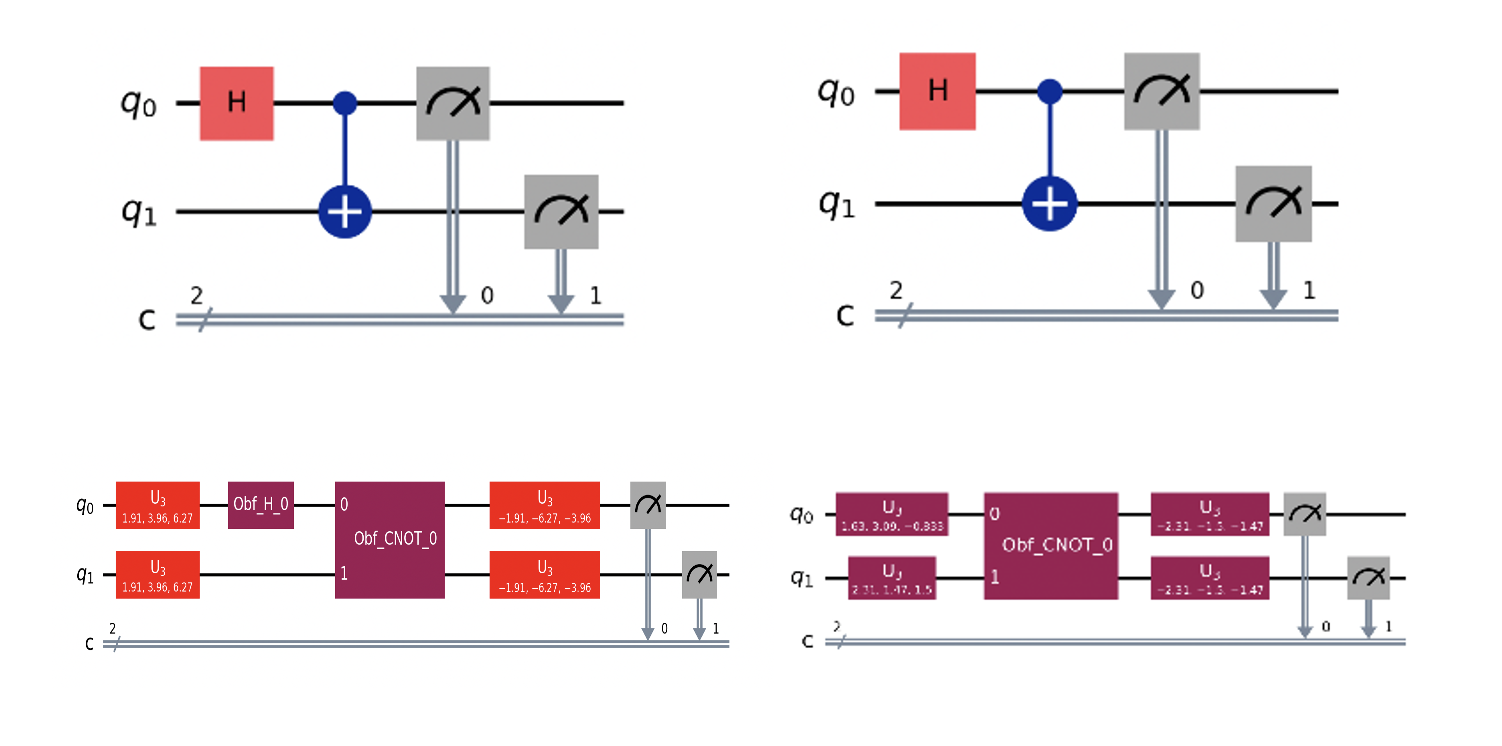} 
                \caption{Illustration of compiler resistance: On the left, the original Bell state circuit and its obfuscated counterpart. On the right, their respective compiled forms. While the original circuit structure is preserved after compilation, the obfuscated circuit appears as a single opaque UnitaryGate, concealing its functionality from quantum compilers and analysis tools}
                \label{fig:compiler_resistance}
            \end{figure}

            A core strength of our \texttt{U3}-conjugation technique lies in its ability to effectively conceal the original gate identities from quantum compilers and reverse-engineering tools. As demonstrated in our QAOA case study (and generally across all obfuscated circuits), when the obfuscated circuit is passed to a transpiler or compiler (such as Qiskit's transpiler), the individual original gates (e.g., Hadamard, CNOT) are no longer recognizable as their standard forms. Instead, they appear as opaque \texttt{UnitaryGate} objects, each representing the complex, randomized matrix resulting from the $U^{\dagger} \cdot G \cdot U$ transformation. Our method's compiler resistance stems from this deliberate lack of identifiable internal structure. Adversaries are forced to treat each transformed gate as an unknown unitary operation, greatly hindering reverse engineering of the proprietary quantum logic. This contrasts sharply with methods that merely insert dummy gates, which are often easily identified and removed by sophisticated optimizers. Fig. \ref{fig:compiler_resistance} visually exemplifies this transformation from a recognizable gate to an opaque block.
                
            \subsection{Overhead Analysis}\label{sec:overhead}
                To make the overhead of the obfuscation explicit, we summarize both the empirical execution-time overhead and the formal structural overhead introduced by the transformation. The execution-time results in Table \ref{tab:eval_metrics_table} and Fig. \ref{fig:time_comparison} show that the runtime increase introduced by obfuscation is extremely small, typically well under one millisecond.
                
                Consider an original circuit with $m$ gates acting on $n$ qubits. For each gate, we insert two single-qubit $U_3$ operations, resulting in an additional
                \[
                G_{m} = 2m
                \]
                single-qubit gates surrounding the original operations.
                
                We further apply $n$ single-qubit $U_3$ gates at the beginning and $n$ at the end of the circuit, contributing:
               \begin{equation*}
                    G_{n} = 2n
                \end{equation*}
                additional gates.
                \\
                Before fusion, the total number of gates is therefore
                \begin{equation*}
                    G_{\text{total, pre-fuse}} = m + 2m + 2n = 3m + 2n
                \end{equation*}
                The $3m$ gates surrounding the original $m$ positions are fused into $m$ effective gates by multiplying the three matrices acting at each position:
                \begin{equation*}
                    3m \longrightarrow m ,
                \end{equation*}
                yielding a final structural gate count of
                \begin{equation*}
                    G_{\text{final}} = m + 2n
                \end{equation*}
                The two global boundary layers contribute a fixed depth increase:
                \begin{equation*}
                    \Delta D = 2
                \end{equation*}
                For example, in the QAOA case study, the computational overhead was negligible because each transformed operation is represented as a single opaque unitary during simulation, contributing no additional internal cost. Because the structural additions are constant and independent of circuit size, the empirical overhead remains negligible even for large workloads.

        \subsection{Results Summary}
\begin{figure}[H]
                \centering
                \includegraphics[width=1\textwidth]{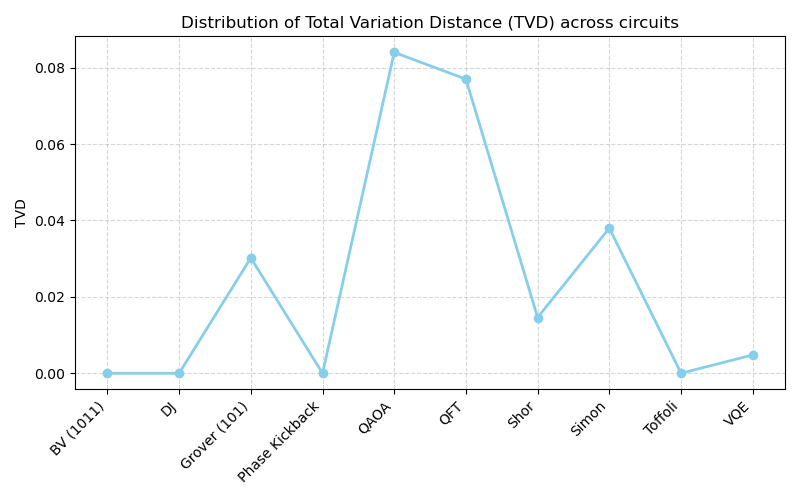}
                \caption{Distribution of Total Variation Distance (TVD) across circuits, with low values indicating strong statistical similarity between original and obfuscated outputs.}
                \label{fig:tvd_distribution}
            \end{figure}

\begin{figure}[H]
                \centering
                \includegraphics[width=0.9\textwidth]{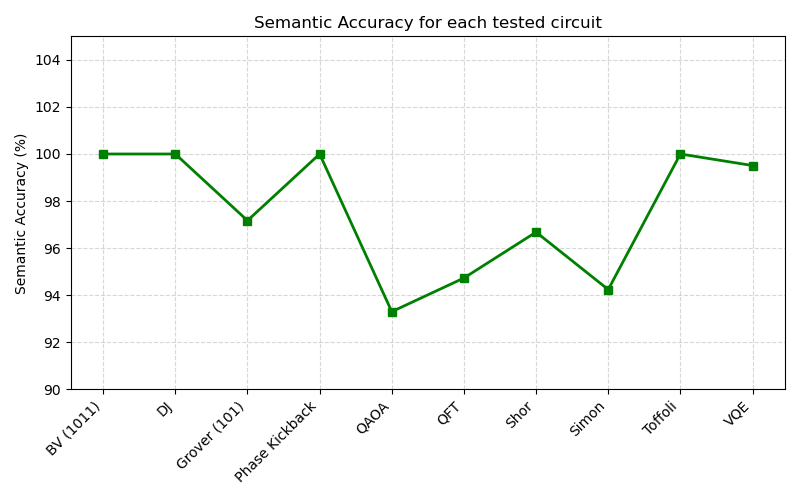}
                \caption{Semantic accuracy for each tested circuit, with high values demonstrating functional correctness of obfuscated versions.}
                \label{fig:accuracy_distribution}
            \end{figure}

\begin{figure}[H]
                \centering
                \includegraphics[width=0.9\textwidth]{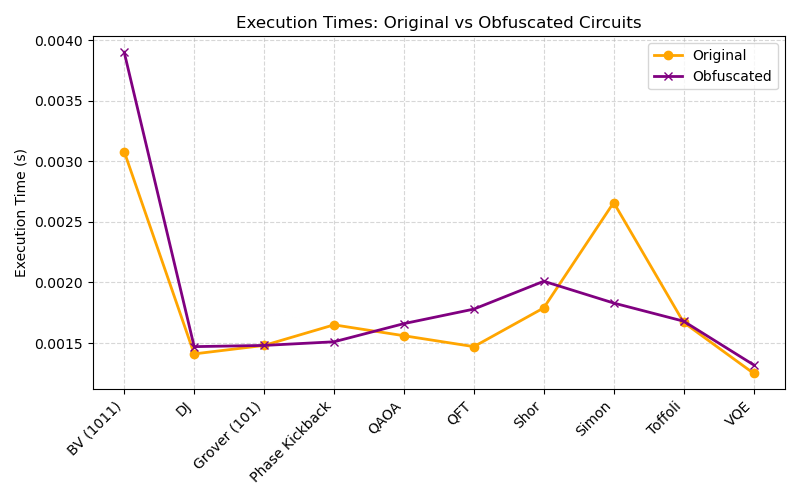}
                \caption{Comparison of execution times between original and obfuscated quantum circuits, showing minimal overhead.}
                \label{fig:time_comparison}
            \end{figure}

            Table~\ref{tab:eval_metrics_table} presents the evaluation metrics for all tested circuits, including QAOA. The results show high semantic accuracy ($>93\%$) and low TVD ($\approx$ 0) across all circuits, confirming that the obfuscation preserves functionality. Execution times remain comparable, with minimal overhead, highlighting the method’s efficiency. Visualizations in Fig. ~\ref{fig:tvd_distribution}, \ref{fig:accuracy_distribution}, and \ref{fig:time_comparison} further illustrate the consistency of TVD, accuracy, and execution times, reinforcing the robustness of the obfuscation technique across diverse quantum algorithms.

           \begin{table}[H]
            \centering
            \caption{Evaluation metrics across quantum circuits.}
            \label{tab:eval_metrics_table}
            \resizebox{0.9\textwidth}{!}{%
            \begin{tabular}{|l|c|c|c|c|}
                \hline
                \textbf{Circuit} & \shortstack{\textbf{Original} \\ \textbf{Time (s)}} & \shortstack{\textbf{Obfuscated} \\ \textbf{Time (s)}} & \shortstack{\textbf{Semantic} \\ \textbf{Accuracy (\%)}} & \textbf{TVD} \\
                \hline
                BV (1011) & 0.00309 & 0.00390 & 100.00 & 0.0000 \\
                DJ & 0.00141 & 0.00147 & 100.00 & 0.0000 \\
                Grover (101) & 0.00148 & 0.00148 & 97.17 & 0.0302 \\
                Phase Kickback & 0.00165 & 0.00151 & 100.00 & 0.0000 \\
                QAOA & 0.00156 & 0.00166 & 93.30 & 0.0840 \\
                QFT & 0.00147 & 0.00178 & 94.73 & 0.0770 \\
                Shor & 0.00179 & 0.00201 & 96.68 & 0.0146 \\
                Simon & 0.00266 & 0.00183 & 94.24 & 0.0380 \\
                Toffoli & 0.00167 & 0.00168 & 100.00 & 0.0000 \\
                VQE & 0.00125 & 0.00132 & 99.51 & 0.0048 \\
                \hline
                \end{tabular}%
            }
            \end{table}

    \section{Security Guarantee}\label{sec:security}
        We quantify the difficulty of deobfuscation under two natural cases:
        
        \subsection{Black-box adversary (no circuit access)}
            In this model, the adversary does not see the obfuscated circuit description. Instead, they have only black-box access to its input-output behavior (e.g., they can prepare input states and observe measurement statistics). Our obfuscation replaces each protected single-qubit gate $G$ by a basis-conjugated form
            \[
                U^{\dagger} G' U,
                \qquad 
                G' = U G U^{\dagger},
            \]
            where the basis rotation $U$ is sampled from the continuous $U3$ parameter domain $(\theta,\phi,\lambda) \in [0,\pi]\times[0,2\pi)\times[0,2\pi)$.
            An adversary who does not see the circuit would need to determine these continuous parameters solely from observed input-output statistics. Because the parameter space is uncountable, the probability of exactly guessing the correct triplet $(\theta,\phi,\lambda)$ approaches $0$.
        
            Under a practical finite-precision discretization of step size $\delta$ (that is, each angle quantized in increments of $\delta$), the number of possible basis rotations is on the order of
            \[
                \left( \frac{2\pi}{\delta} \right)^3
            \]
            and therefore a random-guessing adversary succeeds with probability at most
            \[
                \left( \frac{\delta}{2\pi} \right)^3
            \]
            which is negligible for realistic values of $\delta$. Thus, in the black-box setting, both full and partial obfuscation lead to a difficult continuous parameter-recovery problem.
        
        \subsection{White-box adversary (circuit access).}
            In this stronger model, the adversary has access to the full obfuscated circuit (e.g., QASM or a Qiskit circuit) and can directly observe the explicit gates $U_i^{\dagger}$, $G'_i$, and $U_i$ inserted around protected gates. If all $n$ gates in the circuit are obfuscated, the obfuscation pattern is trivial: the adversary may assume every gate is protected and straightforwardly recover the original gate as
            
            \[
                G_i = U_i^{\dagger} G'_i U_i,
            \]
            thus succeeding with probability $1$.
            
            The problem becomes non-trivial when only a selected subset of gates is obfuscated. Let the original circuit contain $n$ gates, and let the obfuscator randomly choose a hidden subset $S \subseteq \{1,\dots,n\}$ of size $|S|=x$. Only positions in~$S$ are replaced by basis-conjugated blocks $U_i^{\dagger} G'_i U_i$; the remaining gates stay unchanged. A white-box adversary must therefore first determine which gate positions belong to $S$ before applying the algebraic simplification above.
            
            If $S$ is sampled uniformly from all $\binom{n}{x}$ choices of size $x$, then any adversary who attempts to guess the full obfuscation pattern in one shot succeeds with probability
            
            \[
                p_{\mathrm{succ}}(n,x)
                = \Pr[\widehat{S} = S]
                = \frac{1}{\binom{n}{x}}.
            \]
            
            The corresponding min-entropy of the obfuscation pattern is
            
            \[
                H_{\infty}(S)
                = -\log_2 p_{\mathrm{succ}}(n,x)
                = \log_2 \binom{n}{x}.
            \]
            
            This value is $0$ in the trivial cases $x=0$ (no obfuscation) and $x=n$ (all gates obfuscated), but strictly positive for $1 \le x \le n-1$, and maximised when $x \approx n/2$, where
            $H_{\infty}(S) \approx n$ bits. Thus:
    
            \begin{enumerate}
                \item When all gates are obfuscated ($x=n$), a white-box adversary
                can de-obfuscate by simplifying every block.
            
                \item When only a randomly selected subset of gates is obfuscated
                ($1 \le x \le n-1$), the adversary must identify which gates are
                protected; this requires distinguishing among $\binom{n}{x}$
                possible obfuscation patterns. The success probability
                $\dfrac{1}{\binom{n}{x}}$ is exponentially small when $x \approx n/2$.
            \end{enumerate}

    \section{Limitations and Future Work}\label{sec:limitations}
        While the proposed U3-based basis conjugation technique effectively preserves functional correctness and provides strong resistance to compiler-level reverse engineering, it introduces structural overhead that impacts scalability on near-term quantum hardware.

        \subsection{Limitations}
            The obfuscation process can be computationally intensive, especially when applying basis transformations to multi-qubit gates. Constructing these transformations using Kronecker products, as outlined in \cite{Laub2005}, involves manipulating high-dimensional matrices (such as $8 \times 8$ for three-qubit gates), which is feasible for small circuits (e.g., Toffoli). However, for deeper circuits, the large number of gates leads to significant computational costs from repeated matrix operations, limiting scalability \cite{physrev2020}. Generating new random $U3$ parameters for each gate further increases time and memory usage, posing challenges for algorithms with high depth or width.
            
            As analyzed in Section \ref{sec:overhead}, the obfuscation process replaces each protected gate with a conjugated form that introduces additional single-qubit basis rotations, together with two global basis transformation layers applied at the circuit boundaries. Under ideal, noise-free simulation, these additional operations are algebraically fused into opaque unitary blocks, resulting in negligible execution-time overhead and no loss of functional equivalence. Consequently, ideal simulation masks the practical impact of this added structure.
            
            When deployed on NISQ hardware, the additional basis operations introduced by conjugation must be realized as physical gate executions with nonzero error probabilities. Each single-qubit rotation corresponds to a calibrated control pulse whose imperfections accumulate as circuit depth increases, imposing a practical scalability limit even when individual gate fidelities are relatively high. For example, a two-qubit \textsc{CNOT} gate in a depth-$D$ circuit is transformed under basis conjugation into $U^{\dagger}\textsc{CNOT}U$, introducing single-qubit basis rotations on both qubits before and after the entangling operation. While these rotations are algebraically fused into a single opaque unitary during simulation, on physical hardware they must be executed as separate calibrated pulses. Consequently, a circuit containing $m$ such entangling gates incurs $O(m)$ additional single-qubit operations, causing errors to accumulate with depth and limiting scalability for deep circuits such as the Quantum Fourier Transform (QFT) or Variational Quantum Eigensolvers (VQE), despite preserved functional correctness in ideal execution.
            
            Finally, the current implementation does not exploit hardware-specific gate sets or connectivity constraints, which could mitigate some of the depth overhead through optimized decompositions. Incorporating hardware-aware synthesis and calibration-informed transformations remains an important direction for improving scalability on NISQ platforms.

        \subsection{Future Work}
            To overcome these challenges, we plan to pursue several targeted improvements to enhance the technique’s scalability, hardware compatibility, and security. First, to reduce the computational burden of matrix calculations for deep circuits like QFT, or VQE, we will cache precomputed U3-transformed gates (e.g., $U^{\dagger} G U$ for Hadamard or CNOT), minimizing repetitive Kronecker product operations.Second, we will investigate methods to reduce the fidelity losses introduced by the extra single-qubit rotations in the conjugated form $UGU^{\dagger}$, including hardware-calibrated decompositions and approximation-aware synthesis to limit error accumulation. In addition, to better support NISQ devices, we will tailor U3 transformations to match the native gate sets and connectivity of specific quantum hardware, reducing circuit depth and errors. Finally, we also plan to add special gates that trigger errors if tampered with, helping detect unauthorized analysis attempts. These enhancements will make the technique more efficient and secure, supporting applications like protecting intellectual property in cloud-based quantum computing to visualize secure quantum circuits.

    \section{Conclusion}\label{sec:conclusion}
        Our quantum circuit obfuscation method, built on \texttt{U3} gate conjugation, hides the structure of quantum circuits while preserving functionality. By applying randomized basis transformations, it ensures compatibility with \texttt{OPENQASM 2.0} and \texttt{OPENQASM 3.0}. The evaluation presented in this work uses the conservative single-basis variant, while the per-gate randomized variant provides strictly stronger security without altering the functional behavior of the circuit.
        
        We tested the method on a diverse set of circuits---Shor’s algorithm, QAOA, Bernstein-Vazirani, Grover’s algorithm, VQE, Phase Kickback, Deutsch-Jozsa, QFT, Toffoli, and Simon’s algorithm---covering applications from cryptography to quantum chemistry. The obfuscated circuits achieve an average semantic accuracy of at least 93\% and a Total Variation Distance (TVD) below 0.035, indicating minimal changes in output distributions. Using Qiskit’s \texttt{UnitaryGate} to implement transformed gates ($U^{\dagger} G U$) enables seamless integration into quantum circuits, with runtime overhead typically under 1~ms.
        
        These results demonstrate that the method is both practical and versatile, spanning circuits from small Toffoli gates to complex algorithms like QFT and Shor. Future enhancements will focus on improving scalability for larger algorithms, optimizing transformations for NISQ hardware. These advances will strengthen the method’s utility for protecting commercial quantum software, enabling secure cloud execution in quantum computing.
            
    \section*{Declaration of Competing Interests}
        The authors have no competing interests to declare relevant to this article.

    \bibliographystyle{splncs04}
    \bibliography{references}

\end{document}